\documentclass[12pt]{article} \input epsf.tex
\usepackage{graphicx}

\newcommand{\be}{\begin{equation}}
\newcommand{\ee}{\end{equation}}
\newcommand{\bear}{\begin{eqnarray}}
\newcommand{\eear}{\end{eqnarray}}
\newcommand{\ba}{\begin{array}}
\newcommand{\ea}{\end{array}}
\newcommand{\lae}{\begin{array}{c}\,\sim\vspace{-21pt}\\<
\end{array}}
\newcommand{\gae}{\begin{array}{c}\,\sim\vspace{-21pt}\\>
\end{array}}
\newcommand{\CL}{{\cal L}} 
\newcommand{\ord}[1]{\mathcal{O}\hspace{-.5ex}\left(#1\right)}   

\begin{document}

\pagestyle{empty} \begin{titlepage}
\def\thepage {} 

\title{\Large \bf
Nonexotic Neutral Gauge Bosons
}

\author{\normalsize
\bf \hspace*{-.3cm} Thomas Appelquist $^1$,
Bogdan A.~Dobrescu $^2$, Adam R.~Hopper $^1$
\\ \\ 
{\small {\it
$^1$ Department of Physics, Yale University, New Haven, CT 06520, USA }}\\
{\small {\it
$^2$ Theoretical Physics Department, Fermilab, Batavia, IL 60510, USA }}\\
}

\date{ } \maketitle

\vspace*{-7.4cm}
\noindent \makebox[11cm][l]{\small \hspace*{-.2cm} December 4, 2002 (revised: 
May 9, 2003)} 
{\small YCTP-11-02} \\
\makebox[11cm][l]{\small \hspace*{-.2cm} hep-ph/0212073} 
{\small FERMILAB-Pub-02/307-T} \\
{\small } \\

\vspace*{6.9cm}

\begin{abstract}
{\small
We study theoretical and experimental constraints on 
electroweak theories including a new color-singlet and electrically-neutral
gauge boson. We first note that the electric charges of the observed
fermions imply that any such $Z^\prime$ boson may be described by
a gauge theory in which the Abelian gauge groups are the usual hypercharge
along with another $U(1)$ component in a kinetic-diagonal basis. Assuming
that the observed quarks and leptons have generation-independent $U(1)$
charges, and that no new fermions couple to the standard model gauge bosons,
we find that their $U(1)$ charges form a two-parameter family consistent
with anomaly cancellation and viable fermion masses, provided there are
at least three right-handed neutrinos. We then derive bounds
on the $Z^\prime$ mass and couplings imposed by direct production and 
$Z$-pole measurements. 
For generic charge assignments and a gauge coupling of electromagnetic 
strength, the
strongest lower bound on the $Z'$ mass comes from $Z$-pole measurements, and
is of order 1 TeV. If the new $U(1)$ charges are proportional 
to $B - L$, however,
there is no tree-level mixing between the $Z$ and $Z'$, and the best bounds
come from the absence of direct production at LEPII and the Tevatron. 
If the $U(1)$ gauge coupling is one or two orders of
magnitude below the electromagnetic one, these bounds are satisfied for
most values of the $Z'$ mass.}

\end{abstract}

\vfill \end{titlepage}

\baselineskip=18pt \pagestyle{plain} \setcounter{page}{1}

\section{Introduction} \setcounter{equation}{0}

The existence of the gauge bosons
associated with $SU(3)_{C} \times SU(2)_{W} \times U(1)_{Y}$,
together with their measured properties,
comprise some of the most profound information obtained
in high-energy experiments.
Other gauge bosons may exist and
interact with observed matter provided they are sufficiently heavy
or weakly coupled to have escaped detection \cite{Hagiwara:pw}.
New electrically
neutral and color-singlet gauge bosons, usually called $Z'$
bosons, are of special interest. 
They may appear as low-energy manifestations of grand
unified and string theories \cite{Hewett:1988xc},
theories of dynamical electroweak symmetry breaking \cite{Hill:2002ap},
and other theories for physics beyond the standard
model. From a more phenomenological perspective, they
have been hypothesized as explanations for possible discrepancies
between experimental results and standard model predictions
(for two such examples, see \cite{Erler:1999nx,Jenkins:1987ue}).

The extensive literature on $Z'$ bosons often deals with either
$Z'$ couplings arising from particular models,
or with "model-independent" (unconstrained) parametrizations of
the $Z'$ couplings \cite{Leike:1998wr}.
In this paper, we investigate an important 
intermediate situation, in which
the properties of the $Z'$ boson are constrained by generic
conditions on four-dimensional effective field theories that involve
extensions of the standard model gauge symmetry.
This approach leads to a variety of interesting
possibilities for $Z'$ bosons that have not been examined previously.

We start (in Section 2)
by analyzing the most general gauge symmetry that
leads to an additional color-singlet and electrically
neutral gauge boson.
We observe  that the full gauge group may be taken to be $SU(3)_{C}
\times SU(2)_{W} \times U(1)_{Y} \times U(1)_{z}$ where $U(1)_{Y}$
is the usual hypercharge group and $U(1)_{z}$ is an additional
spontaneously broken gauge symmetry. Furthermore, the kinetic
mixing between the $U(1)_{z}$ and $U(1)_{Y}$ gauge fields may,
without loss of generality, be taken to vanish at any particular
scale. This provides a helpful simplification in the analysis of
the effective theory, and distinguishes our approach from
those of \cite{Galison:1983pa} and \cite{Babu:1997st}, where either
an ``off-diagonal'' gauge coupling or a kinetic mixing term are
introduced.

The properties of the $Z'$ boson depend only on the scale of the $U(1)_{z}$
breaking, the $U(1)_{z}$ gauge coupling, and the $U(1)_{z}$ charges of the
various fields. In Section 3, we consider the possible values of these
charges. We restrict attention to the case in which the only fermions
charged under $SU(3)_{C} \times SU(2)_{W} \times U(1)_{Y}$ are the three
generations of quarks and leptons. We also take the $U(1)_{z}$ charges to be
generation independent in order to avoid the constraints from flavor-changing
neutral current processes. Anomaly cancellation in the effective theory then
restricts the $U(1)_{z}$ charges of the standard-model fermions
to depend on at most two free
parameters. The standard-model Yukawa couplings 
determine the $U(1)_{z}$ charge of the Higgs doublet in terms of
these two parameters. We also include a number of right-handed neutrinos, 
which are singlets under $SU(3)_{C} \times SU(2)_{W} \times U(1)_{Y}$, 
and derive the relations among their charges that allow the masses 
required for neutrino oscillations.

In general, there is mass mixing between the $U(1)_{z}$ and
$U(1)_Y$ fields, and this is the origin of the $Z$-pole physics to be
considered here. The tree-level
mixing vanishes only in the case that the $U(1)_{z}$
charges are proportional to $B-L$, because the $U(1)_{z}$ charge of
the Higgs doublet then vanishes. 
There will also be
$U(1)_{z}$-$U(1)_Y$ kinetic mixing generated at the one-loop level 
and above, with its renormalization group scale
dependence. As noted above, however, it may be diagonalized away at any
particular scale.

In Section~\ref{SectionProduction}, we consider the bounds from direct
searches at the Tevatron and LEP on the $Z'$ mass and coupling. This is
particularly interesting if the $U(1)_{z}$ charges are proportional to the
$B-L$ number, so that there is no tree-level mass mixing between the $Z$ and
$Z^\prime$ bosons. When the $U(1)_{z}$ charges are not of the $B-L$ type,
the indirect bounds imposed mainly by the $Z$-pole data are stronger. 
In Sections~\ref{SectionT} and \ref{SectionAnomCoup},
we compute the electroweak oblique and vertex 
corrections, respectively, at
tree level in the effective theory, and determine the current experimental
bounds on the $Z^\prime$ parameters. In Section~\ref{SectionConclusions}, we
summarize our results, and comment on their implications. 

\section{Symmetry breaking pattern}
\setcounter{equation}{0}

A new electrically neutral, color-singlet gauge boson $Z'$ may arise
from various extensions of the standard model gauge group,
including products of a larger number of
semi-simple groups, as well as embeddings of some or all
of $SU(3)_C \times SU(2)_W \times U(1)_Y$ into a larger
group. Any such extended gauge group must have an
$SU(3)_C \times SU(2)_W \times U(1)_1 \times U(1)_2$ subgroup
whose generators are associated with the gluon octet, the
$W^\pm$ bosons, and three $SU(3)_C \times U(1)_{\rm em}$
singlets: the photon, the $Z$ boson, and a $Z^\prime$ boson.

There could be more $Z'$ bosons as well as heavy charged
gauge bosons. The former would require additional $U(1)$ groups,
and would lead to a more complicated mixing pattern.
We assume that any additional $Z'$ bosons are sufficiently
heavy or weakly coupled that they can be integrated out
of the effective theory. Additional charged gauge bosons
would contribute to the mixing of the $Z'$ and $Z$ only
through loops, but they could mix at the tree-level
with the $W$, shifting its mass and contributing to
the $T$ parameter, thus affecting the precision
constraints on the mass and coupling of the $Z'$.
We assume that they, too, are weakly coupled or heavy
enough to be integrated out.

The gauge symmetry must be spontaneously broken to $SU(3)_{C} \times 
U(1)_{em}$.
We take the Higgs sector to consist of a complex doublet field
$H$ and a complex scalar field $\varphi$ [a singlet under $SU(3)_{C} \times 
SU(2)_{W}$],
both of which acquire VEV's. For the purpose of studying the tree-level 
properties of the $Z^{\prime}$, each of these fields may be taken 
to describe either linear or nonlinear
realizations of the (spontaneously broken) symmetry. 
A more elaborate symmetry
breaking sector could be adopted, for example with more 
doublet or singlet scalars, but,
as in the case of additional charged or colored gauge bosons, there would 
be no impact at tree level on the properties of the $Z^{\prime}$. 

We choose a basis for the $U(1)_1 \times U(1)_2$ gauge fields of the
effective theory in which the kinetic terms are diagonal and
canonically
normalized. Because kinetic mixing is induced at the one loop level and
higher, the diagonalization required to do this is scale dependent. For
our purposes, however, since the gauge coupling is assumed to be small
and we are not concerned with energy scales above a few TeV,
the scale dependence is unimportant.

The usual hypercharge gauge group is not in general identified with
either $U(1)_1$ or $U(1)_2$. Upon performing a certain $SO(2)$
transformation on the two gauge fields we can always choose the charge
of $\varphi$ under one of the resulting $U(1)$'s to be zero.
By rescaling this $U(1)$ coupling, we define the corresponding $H$
charge to be +1. We label
this group by $U(1)_Y$ because shortly we will note that the measured
electric charges imply that this is precisely
the standard model hypercharge gauge symmetry. The other
linear combination of
gauge bosons is labelled by $U(1)_z$. The symmetry breaking pattern
requires $\varphi$ to be charged under $U(1)_z$, and we choose
its charge to be +1 by rescaling the $U(1)_z$ gauge coupling.

The mass terms for the three electrically-neutral $SU(2)_W \times
U(1)_Y \times U(1)_z$ gauge bosons, ${W^3}^\mu$, $B_Y^\mu$ and 
$B_z^\mu$, arise
from the kinetic terms for the scalar fields upon replacing $H$ and
$\varphi$ by their VEVs: \be \frac{v_H^2}{8} \left( g {W^3}^\mu - g_Y
B_Y^\mu - z_H g_z B_z^\mu \right) \left( g W^3_\mu - g_Y {B_Y}_\mu - z_H
g_z {B_z}_\mu \right)
+ \frac{v_\varphi^2}{8} g_z^2 B_z^\mu  {B_z}_\mu ~,
\ee where $z_H$ is the $U(1)_z$ charge of $H$, $g_Y$ and $g_z$ are the
$U(1)_Y \times U(1)_z$ gauge couplings, and $v_H$ and $v_\varphi$ are
the VEVs of $H$ and $\varphi$. The properties of $W^\pm$ are not 
affected by the additional $U(1)$, so that $g$ is the usual $SU(2)_W$ 
gauge coupling and $v_H \approx 246$ GeV.

The ensuing mass-square matrix for $B_Y^\mu$, ${W^3}^\mu$, and
$B_z^\mu$
can be written as follows:
\be
{\cal M}^2 = \frac{g^2 v_H^2 }{ 4\cos^2 \theta_w} \; U^\dagger
\left( \ba{ccc} 0 & 0 & 0 \\ 0 & 1 & - z_H t_z \cos\theta_w \\
0 & - z_H t_z \cos\theta_w & \left( r + z_H^2 \right)t_z^2 \cos^2
\theta_w
\ea \right) U ~,
\ee
where $t_z \equiv g_z/g$, $\tan\theta_w = g_Y/g$, $r =
v_\varphi^2/v_H^2$.
The matrix
\be
U = \left( \ba{ccc} \cos\theta_w & \sin\theta_w & 0 \\ - \sin\theta_w &
\cos\theta_w
& 0
\\ 0 & 0 & 1 \ea \right)
\ee relates the neutral gauge bosons to the physical states in the case
$z_H
= 0$.

In general, the relation between the neutral gauge bosons ($ B_Y^\mu,
{W^3}^\mu, B_z^\mu $) and the corresponding mass eigenstates
can be found by diagonalizing ${\cal M}^2$: \be \left( \ba{c} B_Y^\mu
\\
[.2em] {W^3}^\mu
\\  [.1em] B_z^\mu \ea \right) = \left( \ba{ccc} \cos\theta_w & -
\sin\theta_w \cos\theta^\prime & \sin\theta_w \sin\theta^\prime \\
[.15em]
\sin\theta_w & \cos\theta_w \cos\theta^\prime & - \cos\theta_w
\sin\theta^\prime \\ [.15em] 0 &
\sin\theta^\prime & \cos\theta^\prime \ea \right) \left( \ba{c} A^\mu
\\
[.15em] Z^\mu \\  [.15em] Z'^{\mu} \ea \right)
~,
\label{basis} \ee where $Z$ and $Z'$ now denote mass eigenstates and
the angle $-\pi/4\leq \theta^\prime\leq \pi/4$ satisfies 
\be 
\tan 2\theta^\prime
= \frac{2 z_H t_z \cos\theta_w}{(r+ z_H^2) t_z^2 \cos^2\theta_w - 1} ~.
\label{thetaprime}
\ee 
The $Z$ eigenstate, of mass $M_{Z}$, corresponds
to the observed $Z$ boson, while the $Z'$ eigenstate, of mass $M_{Z'}$, 
is the heavy neutral gauge boson not yet discovered. The photon is 
massless, while the $Z$ and $Z'$ masses are given by 
\be M_{Z,Z'} = \frac{g v_H }{
2\cos\theta_w}\left[ \frac{1}{2}\left( (r+z_H^2) t_z^2 \cos^2\theta_w 
+ 1 \right) \mp \frac{z_H t_z \cos\theta_w}{\sin 2\theta^\prime}
\right]^{\! 1/2} ~.
\label{Z-masses} \ee
One can check that $Z'$ is heavier than the observed $Z$ when 
$(r + z_H^2) t_z^2\!\cos^{2}\theta_w > 1$.
In the case where $(r + z_H^2) t_z^2 \cos^{2}\!\theta_w < 1$,
Eq.~(\ref{thetaprime})  
gives $z_H / \sin 2\theta^\prime < 0$, so that $M_{Z'} < M_{Z}$.
This is phenomenologically allowed provided the $Z'$ is
sufficiently weakly
coupled (see Sections~\ref{SectionT} and \ref{SectionAnomCoup}).
Note that $M_{Z'} \rightarrow 0$ in the limit $r\rightarrow 0$.

In the mass-eigenstate basis, Eq.~(\ref{basis}) implies that the piece
of the covariant derivative that contains the photon field may be
written as
\be - i g \sin\theta_w \left(T^3 + \frac{Y}{2}\, \right) A^\mu ~, \ee
where
$T^3$ is the weak-isospin operator and $Y$ is the charge operator
corresponding to the $U(1)_Y$ gauge interaction. Requiring that matter
couples to the photon in the usual way, we are led to the conclusion
that
$g\sin\theta_w$ is the electromagnetic coupling and that $Y$ is the
usual hypercharge operator.

The mass and couplings of the $Z^\prime$ are described
by the following parameters:
the gauge coupling $g_z$, the VEV $v_\varphi$,
the $U(1)_z$ charge of the Higgs, $z_H$,
and the fermion charges under $U(1)_z$ (subject to the
constraints discussed in the next Section).  Note that a
kinetic mixing parameter introduced as in \cite{Babu:1997st},
or an off-diagonal gauge coupling introduced as in \cite{Galison:1983pa},
would be redundant in the framework employed here.

\section{$U(1)_z$ charges}
\setcounter{equation}{0}

We assume that the only fermions charged under $SU(3)_C \times SU(2)_W
\times U(1)_Y  \times U(1)_z$ are three generations of quarks, $q_L^i,
u_R^i, d_R^i$, and leptons, $l_L^i, e_R^i$, $i=1,2,3$, and a number $n$
of right-handed neutrinos, $\nu_R^k$, $k=1, ... , n$, which are singlets
under $SU(3)_C \times SU(2)_W$, and are electrically neutral. 
We label the $U(1)_z$ charges as follows: $z_q, z_u, z_d, z_l, z_e$, 
for the standard model fermions
(assuming a generation-independent charge assignment),
and $z_k$, $k = 1, ... , n$, for the right-handed
neutrinos. In this section we first impose the gauge and the mixed
gravitational-$U(1)_z$
anomaly cancellation conditions to restrict these $U(1)_z $ charges,
and
then study the additional constraints on charges required for the
existence of fermion masses.

\subsection{Anomaly cancellation}

The $[SU(2)_W]^2 U(1)_z$ and $[SU(3)_C]^2 U(1)_z$ anomalies
cancel if and only if
\be
z_l = -3z_q \; \; ; \; \;
z_d = 2 z_q - z_u ~.
\label{first}
\ee
The $[U(1)_Y]^2 U(1)_z$ anomaly cancellation then implies
\be
z_e = -2z_q - z_u ~,
\label{second}
\ee
and the $U(1)_Y [U(1)_z]^2$ anomaly automatically cancels.
Eqs.~(\ref{first}) and (\ref{second}) together lead
to the conclusion that only two independent real parameters,
$z_q$ and $z_u$, 
describe the allowed $U(1)_z$ charges of the quarks
and $U(1)_Y$-charged leptons. Equivalently, 
the $U(1)_z$ charges may be expressed as a linear combination 
of $Y$ and $B-L$: $( z_{u}- z_{q} ) Y + ( 4z_{q} - z_{u} ) ( B - L )$
 \cite{Weinberg:kr}. 
This general
labelling is with respect to our chosen basis in which there is no 
kinetic mixing between the $U(1)$ gauge fields.
The gauge charges of all the fermions and scalars are listed in
Table~\ref{TableCharge}.

\begin{table}[t]
\centering
\renewcommand{\arraystretch}{1.5}
\begin{tabular}{|c| |c|c|c|c|}\hline
& $SU(3)_C$ & $SU(2)_W$ & $U(1)_Y$ & $U(1)_z$ \\ \hline\hline
$q_L^i$ &  3 & 2 & $1/3$ & $z_q$\\ \hline
$u_R^i$ &  3 & 1 & $4/3$ & $z_u$\\ \hline
$d_R^i$ &  3 & 1 & $-2/3$ & $2z_q - z_u$\\ \hline
$l_L^i$ &  1 & 2 & $-1$ & $-3z_q$\\ \hline
$e_R^i$ &  1 & 1 & $-2$ & $-2z_q - z_u$\\ \hline
$\nu_R^k$ , $k=1, ... , n$ &  1 & 1 & 0 & $z_k$\\ \hline\hline
$H$ &  1 & 2 & $+1$ & $-z_q + z_u$\\ \hline
$\varphi$ &  1 & 1 & $0$ & $1$\\ \hline
\end{tabular}

\medskip
\parbox{5.5in}{
\caption{ Fermion and scalar gauge charges. 
\label{TableCharge}}}
\end{table}

Additional restrictions on the $U(1)_z$ charges are imposed by the
mixed gravitational-$U(1)_z$ and $[U(1)_z]^3$
anomaly cancellation conditions. Using Eqs.~(\ref{first}) and (\ref{second}),
these conditions may be written as follows:
\bear
&& \frac{1}{3} \sum_{k=1}^n z_k = -4 z_q + z_u ~, \label{sum} \\
&& \left( \sum_{k=1}^n z_k \right)^{\! 3} = 9 \sum_{k=1}^n z_k^3 ~.
\label{nucharges}
\eear 
Furthermore, the observed atmospheric and 
solar neutrino oscillations require that at least two 
active neutrinos are massive, and that there is flavor mixing,
imposing further restrictions on the 
$U(1)_z$ charges. We address these issues in sections 3.2 and 3.3.

\subsection{Fermion mass constraints}

The $U(1)_z$ charges of the fermions, satisfying the requirement of
anomaly cancellation, allow all the standard model Yukawa 
interactions for the quarks and $U(1)_Y$-charged leptons, provided $H$ 
has charge 
\be z_H = -z_q + z_u ~. 
\label{higgs charge}
\ee 
We impose this condition, as indicated in Table
\ref{TableCharge}, because otherwise only operators of dimension higher
than four could contribute to the quark masses, and it would 
be unlikely that a sufficiently large top-quark mass could be generated.

We next discuss the generation of neutrino masses, required to explain
the current neutrino oscillation data. This will impose further restrictions
on the $U(1)_z$ charges depending on the type of neutrino mass being
generated and the number $n$ of right-handed neutrinos.

Majorana mass terms for the active neutrinos
may be induced by the dimension-five operator
\be
\frac{1}{M} \left(\overline{l^c}_L H \right) \left(l_L
H\right) + {\rm h.c.} ~, \label{mass-l} \ee
provided $4z_q = z_u$ so that $U(1)_z$ invariance is preserved.
In the above operator
the flavor indices are implicit, and $M$ is some mass scale higher than the
electroweak scale.
More generally, when $8z_q - 2z_u$ is an integer,
Majorana neutrino masses can be
induced by a higher dimension version of the operator (\ref{mass-l}),
obtained by including the appropriate power of $\varphi/M$ or
$\varphi^{\dagger}/M$ required by $U(1)_z$ invariance.
With $M$ large enough, these operators can lead naturally to
viable neutrino masses.

If $8z_q - 2z_u$ is non-integer, then the neutrinos can have a
mass spectrum compatible with the atmospheric and solar neutrino oscillation
data if and only if there are at least two right-handed neutrinos present,
and an appropriate Dirac mass operator is $U(1)_z$
invariant.
The dimension-four operator \be \overline{l}_L^i \nu_R^k
\tilde{H} + {\rm h.c.} ~ \label{dirac}, \ee where $i = 1,2,3$, is allowed
providing $z_k = -4 z_q + z_u$.
It may lead to a neutrino mass pattern
consistent with the current data, although a small coefficient is required.
More generally, Dirac neutrino masses are induced when $z_k + 4 z_q - z_u$
is an integer, because there are $U(1)_z$-invariant operators obtained by
multiplying the above Dirac mass terms by 
the appropriate power of $\varphi/M$ or $\varphi^{\dagger}/M$.
A larger absolute value for the $z_k + 4 z_q -
z_u$ integer implies that the dimension of the operators leading to the
Dirac neutrino masses is higher, so that sufficiently small neutrino masses
are generated with larger values for the coefficients.

Finally, right-handed Majorana mass operators of the form \be M^{1-2z_k}
(\varphi^\dagger)^{2z_k} \overline{\nu^c}_R^k \nu_R^{k} + {\rm h.c.} ~
\label{mass-r} \ee are allowed by the $U(1)_z$ invariance if $z_k$ is an
integer or half-integer [for $z_k < 0$, $(\varphi^\dagger)^{2z_k}$ is
replaced by $\varphi^{-2z_k}$]. If the ensuing right-handed neutrino masses
are larger than the electroweak scale, then they can lead to a seesaw
mechanism providing that the corresponding Dirac masses also exist. 
In addition, when a right-handed neutrino participates in
both Dirac and right-handed Majorana mass terms, $U(1)_z$ invariance require
$2(4 z_q - z_u)$ to be an integer, so that left-handed 
Majorana mass terms are also $U(1)_z$ invariant.

Based on the above constraints, we now study the various possibilities
for the fermion charges depending on the number of
right-handed neutrinos, $n$.

\subsection{Fermion charge assignments}

If $n = 0$, then the mixed gravitational-$U(1)_z$ anomaly cancellation
[see Eq.~(\ref{sum})] demands $z_u = 4z_q$. For the trivial
case $z_q = 0$, the only field charged under $U(1)_z$ is $\varphi$.  
If $z_q \neq
0$, the $U(1)_z$ charges of the fermions and the Higgs doublet are
proportional to their hypercharges.
We will refer to this $U(1)_z$ as the ``$Y$-sequential'' 
symmetry.\footnote{The $Y$-sequential $Z^\prime$ does {\it not}
couple to the standard-model
fields with the same couplings as the $Z$. The latter possibility 
has been referred to in the
literature as ``sequential'' ({\it e.g.}, see \cite{Barate:1999qx}), 
but we note that such couplings are not attainable within the field 
theoretic framework employed here.}
In either case, small neutrino masses may be generated
by the operator of Eq.~(\ref{mass-l}). For $n=1$, it follows from
Eq.~(\ref{nucharges}) that $z_1 = 0$, and Eq.~(\ref{sum}) again gives
$z_u = 4z_q$, so that the $U(1)_z$ is either trivial or $Y$-sequential.
Small neutrino masses can be generated by the operators 
of Eq.~(\ref{mass-l}),
as well as the seesaw combination of operators given in
Eqs.~(\ref{dirac}) and (\ref{mass-r}).

For $n = 2$, the anomaly constraint Eq.~(\ref{nucharges}) leads to
the condition $z_1 = -z_2$.
Eq.~(\ref{sum}) then gives $z_u = 4z_q$, as in the $n = 0$ and $n
= 1$ cases, again leading to the trivial
or $Y$-sequential 
possibilities for the $U(1)_z$-charges of the left-handed
neutrinos and the Higgs.
For integer values of $z_1 = - z_2$, all the Dirac and
Majorana masses are allowed as explained in Section 3.2.
For non-integer values of $z_1 = - z_2$,
the left-handed Majorana mass operators
given in Eqs.~(\ref{mass-l}) are allowed,
but the Dirac masses are forbidden. A
$\overline{\nu^c}_R^1 \nu_R^2$ Majorana mass is also allowed,
while the diagonal, right-handed
Majorana mass operators given in Eq.~(\ref{mass-r})
are allowed only if $z_{1,2}$ are half-integers.
Viable neutrino masses are always attainable.

The case $n = 3$ leads to a more general set of possibilities.
The assignment $z_1 = 0$ is similar to the $n=2$ case discussed above,
so that it is sufficient to assume that all three $\nu_R$ charges
are nonzero.
A simple but non-trivial assignment
satisfying the $[U(1)_z]^3$ anomaly cancellation [see
Eq.~(\ref{nucharges})] is $z_1 = z_2 = z_3 \neq 0$. The
condition (\ref{sum}) implies $z_1 = -4 z_q + z_u$,  so that
the Dirac mass operators Eq.~(\ref{dirac}) are $U(1)_z$
invariant. The left-handed Majorana masses
are then allowed only if $z_1$ is an integer or half-integer.
The right-handed Majorana operators (\ref{mass-r})
are also allowed when $z_1$ is an integer or half-integer,
leading to an effective seesaw mechanism
for the neutrino masses.

Other nontrivial assignments for the $z_k$'s are 
also possible with $n = 3$. 
When $z_1=z_2$, the anomaly cancellation conditions,
Eqs.~(\ref{sum}) and especially (\ref{nucharges}), allow a single solution:
\be
z_1 = z_2 = -(4/5)z_3 = -16 z_q + 4 z_u \neq 0
\ee
Viable Dirac neutrino masses  are then allowed when $3(4z_q - z_u)$
is an integer. Left-handed Majorana masses  
are allowed when $2(4z_q - z_u)$ is an integer,
and all types of neutrino masses are allowed
when $4z_q - z_u$ is an integer. For example, in the particular case
$z_1 = z_2 = -4$ and  $z_3=5$, which imposes $4z_q - z_u = 1$,
there are three 
left-handed Majorana masses and two Dirac masses generated by 
dimension-7 operators, a third Dirac mass is generated by 
operators of dimension 12, while 
right-handed Majorana masses are generated by operators 
of dimension ranging from 4 to 13.

There are also solutions with all three $z_k$'s different, 
even when restricted to rational numbers.
For example, the assignment $z_1 = 3$, $z_2 = -17/6$ and  
$z_3=-5/3$, which imposes $4z_q - z_u = 1/2$, allows
left-handed Majorana masses generated by 
dimension-6 operators, no Dirac masses, and a single 
right-handed Majorana mass from a dimension-9 operator.
For $n\ge4$ there are many interesting solutions, such as
$z_1= 4, z_2=z_3= 2, z_4=1$, $4z_q - z_u = -3$, which allows 
three Dirac masses generated by dimension-5 operators,
left-handed Majorana masses generated by 
dimension-11 operators, and right-handed Majorana masses 
generated by operators 
of dimensions ranging from 5 to 11.

Two important conclusions may be drawn from this brief discussion. 
Firstly, the allowed $U(1)_z$ charge assignments of the neutrinos 
permit an array of possible
neutrino mass terms, of both Dirac and Majorana type, 
that can naturally accommodate the
current neutrino oscillation data. Secondly, for $n \geq 3$, the 
left-hand side of the
anomaly condition (3.3) can take on a variety of nonzero values, 
allowing the full, two-parameter family of $U(1)_z$ charges 
for the standard model fermions, as listed in Table \ref{TableCharge}.

\subsection{Some specific models}

Before exploring the phenomenology of the ensuing $Z'$ boson,
we comment on certain restrictions of our two-parameter family of
$U(1)_z$ charges, corresponding to some specific models.
This in turn leads to restrictions on the $\nu_R$ charges.
It is worth recalling at this point that we have adopted a gauge-field
basis
at the outset in which an allowed, dimension-4 kinetic mixing term
between
the $U(1)$'s has been rotated away, and into the $U(1)_z$ charges. If
one
adopts the effective-field-theory attitude that this mixing has arisen
from
some underlying physics and is therefore unknown, then there is no
a-priori
reason to assign any particular values to the $U(1)_z$ charges. On the
other
hand, if they arise from some fundamental theory with small kinetic
mixing,
and if the renormalization group running of the mixing from the
fundamental
scale to that of our effective theory is small, then the values of the
charges might obey certain simple relations as in the following models.

Consider first the case in which the $U(1)_z$ is $U(1)_{B-L}$,
namely the $z$ charges of the standard model fermions are
proportional to their baryon number minus their
lepton number. This corresponds to the restriction
$z_u = z_q$. As we will discuss in Section 4.1, this case
is phenomenologically interesting because the $U(1)_{B-L}$ gauge boson
does not mix at tree level with the standard model neutral gauge
bosons.

Another simple example of  $U(1)_z$ is $U(1)_R$, in which
the $z$ charges are proportional to the eigenvalues of the $T^3$
generator of the global $SU(2)_R$ symmetry (which would be exact in the
limit of equal up- and down-type fermion mass matrices).
In the notation of Table 1, this is the $z_q = 0$ case.

A much studied
example of $Z^\prime$ arises from the left-right symmetric model
after the breaking of the $SU(2)_R$ gauge group \cite{Mohapatra:uf}.
The gauge group is given by
$SU(3)_C \times SU(2)_W \times U(1)_R\times U(1)_{B-L}$.
According to our arguments presented in Section 2,
the $U(1)_R\times U(1)_{B-L}$ product group is equivalent to
$U(1)_Y\times U(1)_z$, where the $U(1)_z$ charges can be
determined by comparing the covariant derivatives
of the two product groups.
The hypercharge gauge coupling imposes a relation
between the $U(1)_R \times U(1)_{B-L}$
gauge couplings, $g_R$ and $g_{B-L}$:
\be
g_{B-L} = \frac{g_Y}{\sqrt{1 - \left(g_Y/g_R\right)^2 }} ~,
\ee
and provides a lower bound for them, $g_R, g_{B-L} > g_Y$.
The $U(1)_z$ charges of the fermions are determined (up to an overall
normalization) in terms of $g_R$:
\bear
\frac{z_u}{z_q} & = & 4 - 3 \frac{g_R^2}{g_{Y}^2} < 1 ~,
\nonumber \\ [.2em]
z_q g_z & = & \frac{g_Y^2}{3 \sqrt{g_R^2 - g_Y^2} } ~.
\eear

\begin{table}[t]
\centering
\renewcommand{\arraystretch}{1.5}
\begin{tabular}{|c|c|c|c|}\hline
$U(1)$ type & label & charge assignment & number of $\nu_R$ \\
\hline\hline
``Trivial'' & $U(1)_0$ & $z_u = z_q = 0$ & any \\ \hline
``$Y$-sequential''  &  $U(1)_Y^\prime$   &  $z_u = 4z_q$ & any \\ 
\hline\hline
``$B-L$'' & $U(1)_{B-L}$ & $z_u = z_q$ & $n \ge 3$ \\ \hline
``Right-handed'' & $U(1)_R$ & $z_q=0$ &  $n \ge 3$ \\ \hline
``Left-right'' & $[U(1)_R\times U(1)_{B-L}]/U(1)_Y$  &
$z_u = z_q(4 - 3g_R^2/g_Y^2)$ &  $n \ge 3$ \\ \hline
``$SO(10)$-GUT'' & $U(1)_\chi$ & $z_u=-z_q$ &  $n \ge 3$ \\ \hline
\end{tabular}

\medskip
\parbox{5.5in}{
\caption{Particular cases of $U(1)_z$ symmetries.
\label{Tablezqzu}}}
\end{table}

Another well-known example of a $Z^\prime$
arises in grand unified theories based on the symmetry breaking pattern
$SO(10) \rightarrow SU(5) \times U(1)_\chi$. The $U(1)_\chi$ charges
of the standard model fermions are given by the $U(1)_z$ charges
when $z_u = -z_q$. There are also $Z^\prime$ bosons studied in the
literature which arise from $U(1)$ gauge group that are non-anomalous
only
in the presence of exotic fermions. An example is provided by the
grand unified theories based on $E_6 \rightarrow SO(10) \times
U(1)_\psi$. Such $U(1)$ gauge groups are not 
included in the two-parameter family
of $U(1)_z$ charges.

The various examples of  $U(1)_z$ groups discussed in sections 3.3 and 3.4
are summarized in  Table~\ref{Tablezqzu}.

\section{Experimental Bounds on the $Z^{\prime}$ Parameters}
\setcounter{equation}{0}

The properties of the $Z^{\prime}$ boson are described primarily
in terms of four parameters:
its gauge coupling $g_z$ (equivalently $t_z \equiv g_{z} / g$), the charges
$z_q$ and $z_u$ (with $z_{\varphi}$ defined to be 1), and the VEV
$v_{\varphi}$ of the singlet field $\varphi$ (equivalently the ratio $ r
= v_{\varphi}^{2} / v_{H}^{2}$).
For example, the mass of the $Z^{\prime}$ is given by
Eq.~(\ref{Z-masses}) in terms of $z_{H}^{2}t_{z}^{2}$ and $rt_{z}^{2}$.
Additional parameters, namely the 
number of right-handed neutrinos and their $z$ charges, are relevant
only if the decay
$Z^{\prime} \rightarrow \nu_R \bar{\nu}_R$ is kinematically open.
We next analyze the bounds set on the four parameters listed above
by the current collider data and fits to the electroweak observables.
For the weak-coupling regime ($t_z \lae 1$) considered here, it is
sufficient to restrict the entire discussion to the tree level.

\subsection{Direct $Z^{\prime}$ production}
\label{SectionProduction}

Direct production provides the best bound on the new parameters if $z_H
(= -z_q + z_u)$ is very small compared with $z_q$ and $z_u$.
This is because when $z_H = 0$, corresponding to
pure $B-L$ coupling, the tree-level mixing of the $B^{\mu}_z$ (Eq.
\ref{thetaprime}) vanishes. The presence of the $Z'$ mass eigenstate
then does not affect the mass or couplings of the $Z$ eigenstate
at tree level, and the constraints from precision $Z$-pole data
on the one-loop mixing of the $Z$ with $Z'$ are rather loose in
the weak-coupling regime ($t_z \lae 1$). We label the $Z'$
by $Z_{B-L}$ in this limit and consider the bounds from its direct
production.

The $U(1)_{B-L}$ charges of the fermions are given by
$z_q=z_u=z_d$ and $z_l = z_e = z_k = -3z_q$, with $k=1,2,3$.
Assuming that the CP-even component of the $\varphi$ scalar
is heavier than $M_{Z^\prime}/2$, and that the
right-handed neutrinos have Majorana masses above
$M_{Z^\prime}/2$, we obtain the following branching fractions
for the $Z_{B-L}$:
${\rm Br}(Z_{B-L} \rightarrow l^+ l^- ) \approx 18/37$,
${\rm Br}(Z_{B-L} \rightarrow {\rm hadrons} ) \approx 10/37$,
${\rm Br}(Z_{B-L} \rightarrow {\rm invisible} ) \approx 9/37$, for
$M_{Z^\prime} < 2 m_t$. These are slightly reduced above 
the $t\bar{t}$ threshold.
If the right-handed neutrinos have Majorana masses below
$M_{Z^\prime}/2$, then the branching fractions listed above
become 9/23, 5/23 and 9/23, respectively.

The LEPII experiments provide direct bounds on any $Z^\prime$ that couples
to $e^+e^-$ and is light enough to be produced. Given that the $Z_{B-L}$ has
larger couplings to the leptons than to the quarks, these direct-production
bounds would appear to be particularly stringent.
We estimate the bound on the gauge coupling $z_l g_z$ above which a
$Z_{B-L}$ of a certain mass would have been detected by the LEPII
experiments.

For the rough estimate of $z_l g_z$ sought here, it is
sufficient to analyze the cleanest decay mode,
$Z_{B-L} \rightarrow \mu^+\mu^-$.
The width of the $Z_{B-L}$ resonance in this channel is
\be
\Gamma(Z_{B-L} \rightarrow \mu^+\mu^-)
\approx \frac{(z_l g_z)^2}{48\pi}  M_{Z^\prime} ~.
\ee
For small $z_l g_z$, the resonance is narrow and hard
to discover. LEPII has run at several center-of-mass energies,
and the bound on $z_l g_z$ is stringent only for values of
$M_{Z^\prime}$ very close to these center-of-mass energies.
To derive this stringent bound we take $M_{Z^\prime}=\sqrt{s}$.
When the width is smaller than the energy spread of the beam,
$\delta E \approx 10^{-3}\sqrt{s}$,
the integrated $Z_{B-L}$ production cross section
is given by \cite{Barger:nn}
\bear
\int d(\sqrt{s}) \sigma(e^+e^- \rightarrow  Z_{B-L} \rightarrow \mu^+\mu^-)
& \approx & \frac{6\pi^2}{M_{Z'}^2} {\rm Br}(Z_{B-L} \rightarrow e^+e^-)
\Gamma(Z_{B-L} \rightarrow \mu^+\mu^-)
\nonumber \\ [.2em]
& \approx & \frac{3\pi(z_l g_z)^2}{148 M_{Z'}} ~,
\label{int}
\eear
where the second line corresponds to a
$Z_{B-L}$ that decays only into standard model fermions
(below the $t\bar{t}$ threshold).
The number $N(Z_{B-L})$ of $\mu^+\mu^-$ events due to the presence of
the $Z_{B-L}$ is then obtained by multiplying Eq.~(\ref{int})
with $L/(2\sqrt{\pi}\delta E)$, where $L$ is the integrated
luminosity at $\sqrt{s} = M_{Z'} \pm \delta E$
(for a review, see \cite{Wu:1984ik}):
\be
N(Z_{B-L}) \approx 1.8 \times 10^{-2}
\frac{\left(z_l g_z\right)^2 L}{M_{Z'}\delta E} ~.
\ee
An additional contribution to $N(Z_{B-L})$ comes from the interference
of the amplitudes for $e^+e^- \rightarrow  Z_{B-L} \rightarrow \mu^+\mu^-$
and $e^+e^- \rightarrow \gamma^*, Z^* \rightarrow \mu^+\mu^-$. However, this
contribution is of order $(1/4 - \sin^2\!\theta_w)^2 M_Z\Gamma_Z L/ M_{Z'}^4$ 
at $\sqrt{s} = M_{Z'}$, and can be neglected in what follows.

The background is mainly due to
$e^+e^- \rightarrow  \gamma^*, Z^* \rightarrow \mu^+\mu^-$, with the number
of events for $M_{Z'} > 100$ GeV well approximated by 
\be
N_{\rm B} \approx \frac{4\pi\alpha^2}{3 s} L 
\left[ 1 + \frac{\tan^4\!\theta_w}{(1-M_Z^2/s)^2}
\left(1 + \frac{1 - 4\sin^2\!\theta_w}{8\sin^4\!\theta_w}\right)^2 \, 
\right] ~.
\ee
At the 95\% confidence level, {\it i.e.},
$N(Z_{B-L}) < 1.96 \sqrt{N_{\rm B}}$,
\be
z_l g_z < 1.3 \times (\delta E)^{1/2}L^{-1/4}
\left(1+ \frac{0.125}{(1-M_Z^2/s)^2}\right)^{\! 1/4} ~.
\ee
The most stringent bound is set by the run at $\sqrt{s} = 188.6$ GeV,
where the combined four LEP experiments accumulated the largest
integrated luminosity, $L \approx 0.7 fb^{-1} \times K$. The
factor $K <1$ takes into account the reduction in the effective
luminosity at $\sqrt{s}$ due to initial state radiation (typically
$K \approx 0.5$).
For $M_{Z'} = 188.6$ GeV,
\be
z_l g_z < 0.7 \times 10^{-3} \left(\frac{0.5}{K}\right)^{1/4}
\left( \frac{\delta E}{0.1 \; {\rm GeV}} \right)^{1/2} ~.
\label{resonant}
\ee
Given that the energy spread $\delta E/E$ at LEP is about $10^{-3}$,
the upper bound on the
$U(1)_{B-L}$ gauge coupling could in principle be set at
two or three orders of magnitude
below the electromagnetic gauge coupling for the particular value of
$M_{Z'}$ where LEP is most sensitive to a narrow resonance.
In practice, however, the searches for narrow resonances at LEP
have been performed by comparing the number of signal versus
background events in energy bins which are much larger than
the energy spread of the beam.
The OPAL, DELPHI and L3 collaborations 
searched for narrow resonances in $e^+e^- \rightarrow \mu^+\mu^-$
at $\sqrt{s} \approx 188.6$ GeV, and set an upper bound on the 
coupling to leptons. Specifically, they have considered 
the $R$-parity violating couplings $\lambda_{131}=\lambda_{232}$
of a tau-sneutrino to $e^+e^-$ and $\mu^+\mu^-$ 
\cite{Abbiendi:1999wm, Abreu:2000ap, Acciarri:2000uh}. 
Although this is a scalar, its impact on the total
cross section can be compared to that of a gauge boson.
The OPAL analysis explicitly included only the total
cross section measurement, so that it can be applied to spin-1
bosons, and therefore 
the limit on $\lambda_{232}$ can be translated into a limit on 
$z_l g_z$.
The main difference between the effects of the sneutrino and the $Z_{B-L}$
on the total cross section
is that the branching ratio of the sneutrino
decay into $\mu^+\mu^-$ considered in 
Ref.~\cite{Abbiendi:1999wm},
${\rm Br}(\tilde{\nu} \rightarrow \mu^+\mu^- ) \approx
\lambda_{232}^2 M_{\tilde{\nu}}/(16\pi \; {\rm GeV})$, 
is much smaller than ${\rm Br}(Z_{B-L} \rightarrow \mu^+\mu^-)$.
Multiplying the  experimental bound $\lambda_{232} < 0.02$ 
by a factor $[{\rm Br}(\tilde{\nu} \rightarrow \mu^+\mu^- ) /
{\rm Br}(Z_{B-L} \rightarrow \mu^+\mu^-)]^{1/2}\approx 9.6\times10^{-2}$,
we find $z_l g_z < 1.9 \times 10^{-3}$ at the 95\% confidence level. 
This is fairly close to the result of our
rough computation, Eq.~(\ref{resonant}), given that the
OPAL analysis corresponds to $\sqrt{\pi}\,\delta E = 2.5$ GeV and a 
luminosity smaller by a factor of four.

For a $Z_{B-L}$ resonance located away from $\sqrt{s}$ the
bound on the couplings is less stringent. To derive the bound in the case
where $M_{Z^\prime} < \sqrt{s}$ we need to take into account initial state
radiation. For our purpose it is sufficient to include the emission of a
single photon by an incoming $e^+$ or $e^-$ \cite{Berends:ie}. The cross
section is given by 
\be 
\sigma \left(e^+e^- \rightarrow \gamma Z_{B-L}
\right) \approx \frac{\alpha}{4\pi} \ln\left(\frac{s}{m_e^2}\right)
\frac{s\Gamma_{Z^\prime}}{M_{Z^\prime}}(z_l g_z)^2
\int_\epsilon^1 dx
\frac{(1/x - 1)\left[ 1 + (1-x)^2 \right]}{\left[ s(1-x)
- M_{Z^\prime}^2\right]^2 + M_{Z^\prime}^2\Gamma_{Z^\prime}^2 } ~,
\ee
where $x$ is the ratio of the photon energy to the beam energy,
and $\epsilon \ll 1$ is an infrared cutoff required to avoid the
soft photon singularity (since the $Z_{B-L}$ resonance studied
here is very narrow and below $\sqrt{s}$,
the $Z_{B-L}$ production is not sensitive to the infrared cutoff).
For a total width $\Gamma_{Z^\prime}$
much smaller than $M_{Z^\prime}$, the above equation yields
\be
\sigma \left(e^+e^- \rightarrow \gamma Z_{B-L} \right)
{\rm Br} \left(Z_{B-L} \rightarrow  \mu^+\mu^- \right)
\approx \frac{3 \alpha}{74} \ln\left(\frac{s}{m_e^2}\right)
(z_l g_z)^2 \frac{s^2 + M_{Z^\prime}^4}{s^2 
\left( s-M_{Z^\prime}^2\right) }
~. \label{cross}
\ee
In the range $M_Z < \sqrt{s} < 2M_Z$,
LEPII has run at the following center-of-mass energies:
$\sqrt{s} \approx 130, 136, 161, 172, 183$ GeV \cite{Blondel:1998qz}.
The corresponding integrated luminosities are about
$3,3,10,10$  and $55 \; pb^{-1}$, respectively. We expect that
large gaps in the sensitivity to a narrow resonance in this
range exist.

For exemplification we consider a $Z_{B-L}$
boson of mass around 140 GeV. 
We first estimate the limit imposed by the run at 
$\sqrt{s} \approx 161$ GeV.
The number $N(Z_{B-L})$ of $\mu^+\mu^-$ events due to the presence of
the $Z_{B-L}$ is obtained by multiplying Eq.~(\ref{cross})
with the combined integrated luminosity accumulated
by the four LEP experiments, roughly $40 \; pb^{-1}$:
\be
N(Z_{B-L}) \approx 3\times 10^4 \left(z_l g_z\right)^2 ~.
\ee
The small effect due to the interference of the
$Z_{B-L}$ and $\gamma^*, Z^* $ amplitudes can again be ignored. 

The background in this case is mainly due to 
$e^+e^- \rightarrow  \gamma^*\gamma, Z^*\gamma \rightarrow \mu^+\mu^-\gamma$.
The number of $\mu^+\mu^-$ background events in an energy bin of size 
$\Delta E$, at the reduced center-of-mass energy of 
$\sqrt{s^\prime} = 140$ GeV, is given approximately by 
\bear 
N_B & \approx & 
\frac{2\alpha^3 L \, \Delta E}{3 M_{Z^\prime}}
\left(1+ \frac{0.125}{(1-M_Z^2/M_{Z^\prime}^2)^2}\right)
\ln\left(\frac{s}{m_e^2}\right)
\frac{s^2 + M_{Z^\prime}^4}{s^2 \left( s-M_{Z^\prime}^2\right) }
\nonumber \\ [.2em]
& \approx & 6.4 \times \left(\frac{\Delta E}{5 \; {\rm GeV}}\right) ~.
\eear
At the 95\% confidence-level we find
\be
z_l g_z < 1.2 \times 10^{-2} 
\left(\frac{\Delta E}{5 \; {\rm GeV}}\right)^{\! 1/4}~.
\label{lep-bound}
\ee
Summing all the events for the runs at 
$\sqrt{s} \approx 161, 172, 183, 188.6$ GeV, as well as those between
192 and 208 GeV where an integrated luminosity of about $1.7 fb^{-1}$ has
been accumulated by the four experiments, increases both $N(Z_{B-L})$ 
and $N_B$ by a factor of 19.3, so that the upper bound on $z_l g_z$ 
decreases only by a factor of two.
Thus, for a $Z_{B-L}$ resonance with $M_{Z^\prime} \approx 140$ GeV
the upper bound on the
$U(1)_{B-L}$ gauge coupling that could be set by a combined analysis 
of all LEP data is a factor of 50
below the electromagnetic gauge coupling.
This can again be compared with  the OPAL limit on 
the $R$-parity violating couplings,  
$\lambda_{131}=\lambda_{232} < 0.07$
for a tau-sneutrino mass of 140 GeV, which is based on 
an analysis of data up to $\sqrt{s} = 189$ GeV 
\cite{Abbiendi:1999wm}.
Multiplying by a factor of 
$[{\rm Br}(\tilde{\nu} \rightarrow \mu^+\mu^- ) /
{\rm Br}(Z_{B-L} \rightarrow \mu^+\mu^-)]^{1/2}$
we find $z_l g_z < 2.0 \times 10^{-2}$ at the 95\% confidence level. 

The conclusion so far is that the experimentally allowed region
in the $z_l g_z$ versus $M_{Z^\prime}$ plane has a
complicated shape, with the upper bound on $z_l g_z$ varying
between $10^{-2}$ and $10^{-3}$ for $M_Z \lae M_{Z^\prime}
 \lae 210 \; {\rm GeV}$. We do not expect that a more refined analysis,
including other decay modes, other observables, and numerical simulations,
would change this conclusion. 
We recommend that the LEP collaborations
analyze their data in search of a narrow $Z^\prime$ resonance, either
discovering a signal or deriving a precise exclusion region in the $z_l g_z
- M_{Z^\prime}$ plane.

For 10 GeV $\lae M_{Z^\prime} \lae M_Z$ the situation does not change
qualitatively, but for lighter $Z_{B-L}$ bosons more stringent
limits on $z_l g_z$ can be placed using measurements of
rare meson decays and various other data \cite{Carlson:1986cu}.
We will not discuss further the
case of a very weakly coupled  $Z_{B-L}$ boson.

When the $Z_{B-L}$ boson is heavier than the highest LEP center-of-mass
energy of 209 GeV, the sensitivity of the LEP experiments to a narrow
$Z_{B-L}$ resonance decreases significantly. We first estimate the lower
bound on $M_{Z^\prime}$ for a $g_z z_l$ equal to the electromagnetic
coupling, $e \approx 0.3$, by adapting the bounds set by the ALEPH
and DELPHI Collaborations on various $Z^\prime$ gauge bosons 
\cite{Barate:1999qx, Abreu:2000ap}. Given
that the $Z_{B-L}$ boson couples with the same strength to left- and
right-handed fermions, there are no corrections to the forward-backward
asymmetries. The large leptonic branching fraction imply that the best 
LEPII limits on $Z_{B-L}$ come from the measurement 
of $\sigma(e^+ e^- \rightarrow
Z_{B-L} \rightarrow \mu^+ \mu^-, \tau^+ \tau^-)$.
The analyses in Ref.~\cite{Barate:1999qx, Abreu:2000ap} focusing on the 
$Z^\prime \equiv Z_\psi$
associated with the $U(1)_\psi$ subgroup of $E_6$ in grand unified theories
are well suited for application to our $Z_{B-L}$ because both bosons do not
induce forward-backward asymmetries. (This is true
in the case of the $Z_\psi$ given that its squared-couplings to all quarks 
and leptons are equal.)
Using the normalization for the $U(1)_\psi$ coupling prescribed in
Ref.~\cite{Barate:1999qx}, in the case of 
$U(1)_{B-L}$ we find $M_{Z^\prime} \gae 300$
GeV for $g_z z_l \approx e$. For a coupling to fermions weaker than
the electromagnetic one, this limit is relaxed.

We now turn to the limits in the $M_Z^\prime$ versus 
$g_z z_l$ set by the CDF and D0 experiments using
data obtained in the Run I at the Tevatron.
The data is analyzed such that an exclusion plot in the 
$\sigma(p\bar{p} \rightarrow Z^\prime) \times 
{\rm Br}(Z^\prime \rightarrow \mu^+\mu^-, e^+ e^-)$ versus
$M_Z^\prime$ plane is obtained \cite{Abe:1997fd, Abachi:1996ud}.
The theoretical curve in this plane in the case of the $Z_{B-L}$
may be derived by comparing again with the case of $Z_\psi$, analyzed in 
Ref.~\cite{Abe:1991vn,Abe:1994ns}.
Assuming that the $Z_\psi$ may decay only into standard model 
fermions we find ${\rm Br}(Z_\psi \rightarrow \mu^+\mu^-, e^+ e^-) = 1/12$,
which is smaller than the same quantity for the $Z_{B-L}$ by a 
factor of 37/144. Multiplying this quantity by the 
squared ratio of the $Z_{B-L}$ and $Z_\psi$ couplings to quarks 
we obtain
\be
\frac{\sigma(p\bar{p} \rightarrow Z_{B-L}) \times 
{\rm Br}(Z_{B-L} \rightarrow \mu^+\mu^-, e^+ e^-)}
{\sigma(p\bar{p} \rightarrow Z_\psi) \times 
{\rm Br}(Z_\psi \rightarrow \mu^+\mu^-, e^+ e^-)} =
\frac{16}{37} \left( \frac{g_z z_l}{e} \right)^2 ~.
\ee
We derive bounds on the $Z_{B-L}$ mass and couplings 
by comparing the theoretical curve given in Fig.~3.a 
of Ref.~\cite{Abe:1994ns}, shifted by the above ratio,
with the 95\% C.L. upper limit set by the CDF Collaboration 
(Fig.~3 of Ref.~\cite{Abe:1997fd}).
For $M_{Z^\prime}$ in the range that is kinematically accessible 
at LEP, the bound on $g_z z_l$ set by LEP [see Eq.(\ref{lep-bound})] 
makes $\sigma(p\bar{p} \rightarrow Z_{B-L}) \times 
{\rm Br}(Z_{B-L} \rightarrow \mu^+\mu^-, e^+ e^-)$
about three orders of magnitude smaller than the limit set 
at the Tevatron. For higher masses, however, the Tevatron bounds 
become more stringent than the LEP bounds. 
We estimate $M_{Z^\prime} > 480$ GeV for $g_z z_l = e$.

\subsection{Weak-isospin breaking }
\label{SectionT}

We next return to the general case in which the Higgs $U(1)_z$ 
charge, $z_H$, is
non-negligible and consider bounds arising from data at the $Z$
pole. Tree-level mass mixing now leads to modifications to both
the mass and couplings of the $Z$. We begin with the mass shift,
expressing the result in terms of the $T$ parameter. Having chosen
to use a basis in which there is no kinetic mixing between the
$U(1)$ gauge fields, there are no tree-level contributions to the
$S$ or $U$ parameters.

The $T$ parameter for our effective theory is \be \alpha T =
-\frac{\Pi_{ZZ}^{new}}{M_Z^2} ~, \ee where $\alpha$ is the fine
structure constant evaluated at the $Z$ mass, 
and $\Pi_{ZZ}^{new} = M_Z^2 -
g^2v_H^2/(4\cos^2\theta_w) $ is the contribution of new physics to
the self-energy of the $Z$. The new, tree-level physics here
arises only from mass mixing and is therefore scale independent.
(At tree level the $W$ mass is unaffected by the addition of
$U(1)_z$, and so makes no contribution to the $T$ parameter.)
Using the expression for $M_Z^2$ given in
Eq.~(\ref{Z-masses}), we have  
\be 
\frac{\alpha T}{1+\alpha T} = 
\frac{1}{2} \left[1-(r+z_H^2)t_z^2\cos^2\!\theta_w \right] +
\frac{z_Ht_z\cos\theta_w}{\sin2\theta'}~, 
\ee 
where $\theta'$ is given by Eq.~(\ref{thetaprime}).

The dependence of $T$ on $r$ and $\theta'$ can be re-expressed as
a dependence on $M_Z^2/M_{Z'}^2$ and $z_Ht_z$ by making use of
Eqs.~(\ref{thetaprime}) and (\ref{Z-masses}). The result is \be
\frac{\alpha T}{1+\alpha T} = \frac{1}{2} \left[ 1-
\frac{M_{Z}^2}{M_{Z'}^2} \mp \sqrt{\left(1-
\frac{M_{Z}^2}{M_{Z'}^2}\right)^2 - 4z_H^2t_z^2 \cos^2\theta_w
\frac{M_{Z}^2}{M_{Z'}^2}} \,\right],~\label{delrho} 
\ee 
where one takes the upper (lower) sign if $M_Z$ is less
(greater) than $M_{Z'}$. Notice that $T$ now depends on only two
new parameters: the mass of the $Z'$ boson and the magnitude of
the coupling strength of the $Z'$ boson to the Higgs. The reality
of $T$ (the positive semi-definiteness of the quantity in the
square root) follows from Eq.~(\ref{Z-masses}).
Examination of Eq.~(\ref{delrho}) leads one to conclude that $T$ is
greater (lesser) than zero for $M_Z$ lesser (greater) than $M_{Z'}$.

\begin{figure}[t]
\begin{center}
\includegraphics*[1.5in,6.7in][6.7in,9.5in]{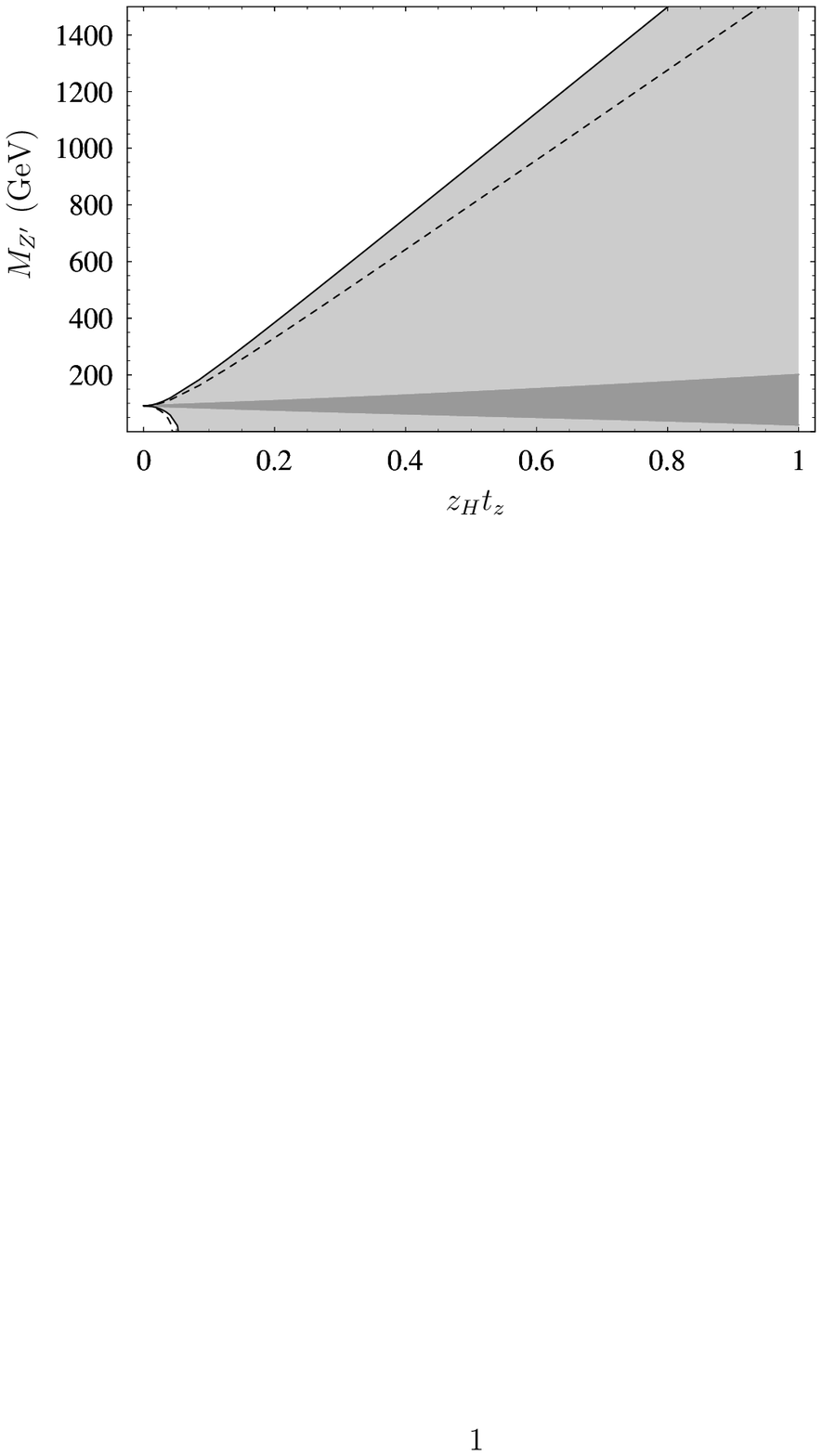}
\caption{Bounds at the $95\%$ confidence level in the
$(z_Ht_z,M_{Z'})$ plane, where $z_H t_z= z_H g_z/g$ is the ratio of
the $U(1)_z$-coupling of the Higgs doublet to the weak coupling.
Electromagnetic strength corresponds to $z_H t_z\simeq
0.48$. The limit $z_H t_z \rightarrow 0$  represents either an extremely 
weakly coupled new boson, or a model with $z\propto B-L$. 
The disallowed region due
to the bound on the $T$ parameter for $M_H = 115$ GeV is shaded light
gray. The dotted line shows the shift in the bound for $M_H = 300$
GeV. Here we have used $T = -0.02\pm 0.13 (+0.09)$, where the
central value corresponds to $M_H = 115$ GeV and the shift (in
parentheses) to $M_H = 300$ GeV~\cite{Hagiwara:pw}. 
\label{FigureT}}
\end{center}
\end{figure}

Experiment demands that $\alpha T \ll 1$. This is assured, for
example, if $M_Z^2/M_{Z'}^2 \ll 1$ or if $z_H^2t_z^2 \ll 1$. More
generally, we plot in Fig.~\ref{FigureT} the allowed region in
the $(z_Ht_z,M_{Z'})$ parameter space due to current constraints on
$T$. The horizontal axis corresponds to the strength of the $z$
coupling of the Higgs, $z_{H}g_z$, ranging from extremely weak up
to roughly twice electromagnetic strength. Bounds are symmetric about
$z_Ht_z = 0$; bounds for negative values of $z_Ht_z$ are not
shown. 
The region allowed by Eq.~(\ref{Z-masses}),
\bear
M_{Z'} & > & M_Z \left(z_Ht_zc_w + \sqrt{1+z_H^2t_z^2c_w^2} \right) 
\;\; , \;\; {\rm for} \; M_{Z'} > M_Z ~,
\nonumber \\ [.2em]
M_{Z'} & < & M_Z \sqrt{\frac{1-|z_H| t_z c_w}{1+|z_H| t_z c_w} }
\;\; , \;\; {\rm for} \; M_{Z'} < M_Z ~,
\eear
is outside the darkly shaded area.
The region allowed by current
constraints on $T$, explained in the figure caption, is outside
the lightly shaded area. Several features are worthy of note. The
lower limit on $M_{Z'}$ is approximately $0.9$ TeV (at the $95\%$
confidence level) with $z_{H}g_z$ of electromagnetic strength
($z_H g_z = e$, $z_Ht_z = \sin\theta_w \approx 0.48$),
while the bound weakens
significantly for smaller $z_Ht_z$. The $Z'$ can actually be lighter
than the $Z$, but this requires very small values for $z_{H}t_z$.
We note that varying $M_H$ from $115$ to $300$ GeV leads to
roughly a $15\%$ shift in the bound for $M_{Z'}$ for a given
$z_Ht_z$.

\setlength{\unitlength}{1in}
\setlength{\unitlength}{1pt}

Finally, we observe that a simple relation exists among the mixing
angle $\theta'$ (given by Eq.~(\ref{thetaprime})), $M_{Z'}$, and
$T$: \be \alpha T = \sin^2\theta'\left(\frac{M_{Z'}^2}{M_Z^2}-1
\right).~\label{Tandthetaprime}\ee Note that if the Higgs doublet
is uncharged under $U(1)_z$ (and so $T=0$), then $\theta'=0$ and
the $Z$ has no anomalous couplings at tree level. From this
expression, it can be seen that through most of the region of
Fig.~\ref{FigureT} allowed by the experimental constraints on $T$,
\,$\theta' \ll 1$.

\subsection{Anomalous $Z$ couplings}
\label{SectionAnomCoup}

We next analyze the tree-level couplings of the neutral gauge bosons to
matter. In terms of the mass eigenstates, the interaction takes the form \be
\CL_{Z,Z'} = \frac{g}{\cos\theta_w} \sum _{f}\overline{f} ~ \gamma^{\mu}
\left(c_f \; , \;  c'_f \right)
\left(\begin{array}{cc}
\cos\theta' & -\sin\theta' \\
\sin\theta' & \cos\theta' \\
\end{array}\right)
\left( \begin{array}{c} Z_\mu \\ Z'_\mu \\ \end{array} \right)f ~, 
\label{mattercoupling} 
\ee 
where $f$ ranges over the chiral fields
$u^i_L$, $d^i_L$, $u^i_R$, $d^i_R$, $\nu^i_L$, $e^i_L$, $e^i_R$, and
$\nu^k_R$. The couplings $c_f$ and $c'_f$ are given by 
\bear c_f & = &
T^3_f-q_f\sin^2\theta_w \nonumber \\ 
c'_f & = & z_ft_z\cos\theta_w/2, 
\eear
where $Y_f$ and $z_f$ are the hypercharge and $U(1)_z$ charge 
of fermion $f$ (see Table~\ref{TableCharge}), 
while $T^3_f$ and $q_f$ are its weak isospin and 
electric charge, which satisfy $q_f = T^3_f + Y_f/2$.
Since the couplings of the $Z$ to matter are known so 
precisely, the constraint
$\theta^\prime\ll 1$ must hold.

In the standard model, electroweak physics is conveniently described in
terms of the electromagnetic coupling, the Fermi constant, 
and the physical $Z$ mass
(in addition to the particle masses and CKM matrix elements). Following this
convention, we focus on $\CL_Z$ from Eq.~(\ref{mattercoupling}) and
reexpress it in terms of these parameters by way of defining a new, physical
weak angle $\theta_W$ such that \be \frac{G_F}{\sqrt2} \equiv \frac{\pi
\alpha(M_Z)}{2\sin^2\theta_W\cos^2\theta_W M_Z^2}~,\ee where
$\alpha(M_Z)$ is the fine-structure constant defined at the $Z$ mass [$\alpha(M_Z)^{-1} = 128.92\pm 0.03$]. For our
effective theory at tree level, the relation between $\theta_w$ and
$\theta_W$ is given by \be \sin^2\theta_W\cos^2\theta_W =
\sin^2\theta_w\cos^2\theta_w(1+\alpha T)~.\ee

Keeping terms to first order in $\alpha T$ and to order $\theta'^2$, the
interaction of the $Z$ with matter may now be written as 
\be \CL_Z =
\frac{e}{\sin\theta_W\cos\theta_W}\left[1+\frac{1}{2}(\alpha T-
\theta'^2)\right] \sum_f\bar{f} \gamma^\mu(g_f+\delta g_f)f
Z_\mu ~,
\label{gsubf} 
\ee 
where terms of order $\alpha T\,\delta g_f$ and  $\theta'^2\,\delta g_f$ 
are discarded. Here $e$ is the 
electromagnetic gauge coupling,
and $g_f$ and $\delta g_f$ are given by
\bear 
g_f &=& T^3_f - q_f \sin^2\theta_W \nonumber \\
\delta g_f &=& q_f\alpha T
\frac{\sin^2\theta_W\cos^2\theta_W}{\cos^2\theta_W -\sin^2\theta_W} +
z_ft_z\cos\theta_W\frac{\theta'}{2}~ \label{deltag}.\eear In the limit
$M_Z\ll M_{Z'}$ the term of order $\theta'^2$ is suppressed relative to the
others, but in general, and particularly for $Z'$ very light compared to
$Z$, it can happen that $\theta'^2\sim \alpha T$ [see
Eq.~(\ref{Tandthetaprime})]. 

Using Eqs.~(\ref{gsubf}) and (\ref{deltag}), any measurable quantity
depending on the $Z$-pole coupling to matter may be expressed (at
tree-level) in terms of $M_Z/M_{Z'}$ and two couplings which we take to be
$z_qt_z$ and $z_Ht_z$. Having expressed $\CL_Z$ in terms of the physical
weak angle $\theta_W$, the prediction for the new theory will
be the radiatively-corrected standard model prediction, plus a small shift
due to new physics that depends on the parameters $M_{Z'}$, $z_Ht_z$, and
$z_qt_z$.

A precise bound in the $Z^\prime$ parameter space can be 
obtained by performing a global fit to all the electroweak data.
However, in order to understand the dependence of the observables
on the $Z^\prime$ parameters, we restrict attention here to 
two well-measured,
representative $Z$-pole observables, in addition to the $T$ parameter:
the total decay width of the $Z$ boson, $\Gamma_Z$, and the
left-right asymmetry of the electron.
We expect that the bounds derived this way will not 
be substantially different than those set by a global fit.

\begin{figure}[t]
\begin{center}
\includegraphics*[1.5in,6.7in][6.7in,9.5in]{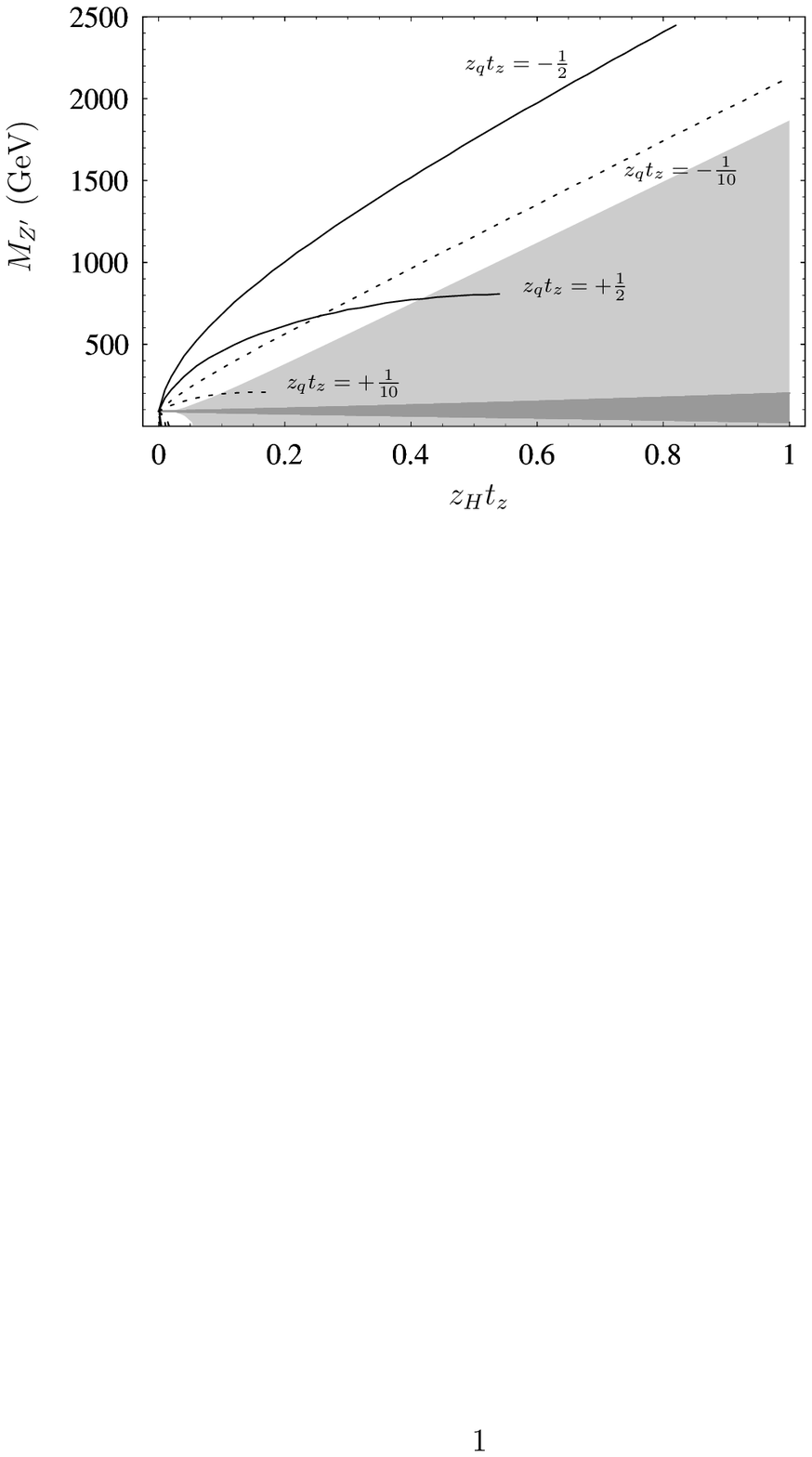}
\caption{Bounds from precision measurements of the total $Z$ width at the
$95\%$ confidence level plotted as contours in the $(z_Ht_z,M_{Z'})$ plane,
for $z_qt_z = \pm\frac{1}{10},\pm\frac{1}{2}$. In each case the allowed region is to the left of  
the line. Included in gray are bounds
from $T$ for $M_H=115$ GeV. ~\label{figzwidth} }
\end{center}
\end{figure}

The current experimental
value of $\Gamma_Z$, $2.4952\pm 0.0023$ GeV, is in excellent 
agreement with the standard
model prediction of $2.4966\pm 0.0016$ GeV~\cite{Hagiwara:pw}. 
The change in the $Z$ couplings due to the presence of the
$Z^\prime$ boson leads to a shift in $\Gamma_Z$,
resulting in the bound shown in Fig.~\ref{figzwidth}. Since we are
interested in comparing bounds on the parameters $M_{Z'}$, $z_qt_z$, and
$z_Ht_z$ to those from the $T$ parameter, we show bounds for a given value
of $z_qt_z$ as contours in the $(z_Ht_z,M_{Z'})$ plane. To understand the
qualitative features of Fig.~\ref{figzwidth}, it is helpful to consider the
limit $M_Z'\gg M_{Z}$, which is reliable over much of the vertical range, in
which case $\Gamma_Z$ takes the simple form 
    \be \Gamma_Z =
    \Gamma_Z^{\it SM}\left[1+(0.55 z_H^2 - 1.29 z_H z_q) t_z^2 M_Z^2/M_{Z'}^2
    +\ord{M_Z^4/M_{Z'}^4}\right], \label{zwidth}\ee 
where $\Gamma_Z^{\it SM}$ is the
current standard model value for the total width. For $z_H$ large relative
to $z_q$ (the $U(1)_R$ limit), Eq.~(\ref{zwidth}) implies that the bound on
$M_{Z'}$ grows linearly with $z_Ht_z$, with a slope of roughly $2$ TeV per
unit $z_Ht_z$. In the limit where $z_q$ is large relative to $z_H$ (the
$z\propto B-L$ limit) the bound grows as $\sqrt{z_Ht_z}$. These features
are reflected in Fig.~\ref{figzwidth}. We note that for $z_Ht_z\ll 1$
there is an allowed region of parameter space with $M_{Z'}<M_Z$.

Fig.~\ref{figzwidth} shows that the bounds from $\Gamma_Z$ depend on the
magnitude of the coupling $z_{q}t_z$, as well as its sign relative to
$z_Ht_z$ which we have taken to be positive. For $z_Ht_z$ and $z_qt_z$ of
electromagnetic strength, {\it i.e.}, $z_Ht_z=\sin\theta_w$, $z_q t_z = \pm \sin\theta_w/3$,
the $T$-parameter bound is $M_{Z'}>0.9$
TeV, while the $\Gamma_Z$-bound is significantly weaker if $z_Ht_z$ and
$z_qt_z$ are of the same sign. If they are of opposite sign the bound from
$\Gamma_Z$ is stronger than that from $T$, with $M_{Z'}>1.2$ TeV. For 
sufficiently small
$z_Ht_z$ the $\Gamma_Z$-bound is always stronger than the $T$-parameter bound
(and both vanish at tree level in the $z_Ht_z\rightarrow 0$ limit).

\setlength{\unitlength}{1in}
\setlength{\unitlength}{1pt}

We have also examined bounds coming from the left-right asymmetry of the
electron, $A_e$.
The experimental value, from the angular distribution of
the $\tau$ polarization~\cite{Hagiwara:pw}, is $0.1498\pm 0.0048$; the
standard model predicts $0.1478\pm 0.0012$. 
An expression analogous to Eq.~(\ref{zwidth}) can be derived for $A_e$, 
    \be A_e =
    A_e^{\it SM}\left[1+(24z_H^2+60z_Hz_q)t_z^2M_Z^2/M_{Z'}^2
    +\ord{M_Z^4/M_{Z'}^4}\right]. \label{ae}\ee 
We find that the bounds from $A_e$ are complementary to those from
$\Gamma_Z$: the former are comparable with the latter for opposite signs
of $z_qt_z$ (this fact is suggested by Eqs.~(\ref{zwidth}) and
(\ref{ae})).
For $z_Ht_z$ and $z_qt_z$ of electromagnetic strength, 
and of the same sign, one finds that the bound from $A_e$ 
is $M_{Z'}>1.0$ TeV at the 95\% confidence level.  

In summary, $M_{Z'}> {\cal O}(1)$ TeV in
models with $z_H t_z$ and $z_qt_z$
of electromagnetic strength.

\section{Conclusions}
\label{SectionConclusions}

Our study of new color-singlet and electrically neutral gauge bosons
shows that there remain many possibilities to explore even in 
the simplest extensions of the standard model.
We have first demonstrated that the
$SU(3)_{C} \times SU(2)_{W} \times U(1)_a \times U(1)_b$
gauge group is consistent with the measured electric charges
of the observed fermions only if it is equivalent to
the $SU(3)_{C} \times SU(2)_{W} \times U(1)_Y \times U(1)_z$
gauge group, where $U(1)_Y$ is the standard model hypercharge,
and $U(1)_z$ is a new gauge group in a kinetically-diagonal
basis.
Symmetry breaking is described, without loss of generality
for the purposes of this paper, by the usual doublet 
Higgs along with a single complex scalar, whose $U(1)_z$
charge is +1.

We have then concentrated on the case where 
the $U(1)_z$ symmetry is non-anomalous, 
the $U(1)_z$ charges of the observed fermions are 
generation independent, and any new fermions are singlets 
under $SU(3)_{C} \times SU(2)_{W} \times U(1)_Y$.
We allow for an arbitrary number of right-handed 
neutrinos charged under $U(1)_z$. 
As long as there are at least three right-handed 
neutrinos, a continuous family of $U(1)_z$-charge 
assignments is consistent with both anomaly cancellation 
and the existence of fermion masses.
The $U(1)_z$-charges of the observed fermions 
depend on two parameters, chosen to be the charges $z_q$ and 
$z_u$ of the left-handed quark doublets and  
right-handed up-type quarks. 
The $U(1)_z$ charge of the Higgs doublet has to be given by
$z_H = z_u - z_q$ in order to allow the existence of the 
top Yukawa coupling. Although the $U(1)_z$ charges of the 
right-handed neutrinos are not uniquely determined,
the anomaly cancellation conditions allow only a limited set
of charge assignments. Moreover, each of these right-handed neutrino
charge assignments 
implies a different set of higher-dimensional operators
that could generate the neutrino masses. As a byproduct, it
is possible to obtain viable neutrino mass matrices even when all the
higher-dimensional operators have coefficients of order unity.

Some fermion charge assignments correspond to relatively 
simple relations between $z_q$ and $z_u$.
Among them are several of 
the popular models in the literature, as well as
other simple charge assignments
that, to the best of our knowledge, have not been analyzed
before.
In fact, in a general effective field theory arising 
from unspecified underlying dynamics, and involving 
renormalization-group running from 
the fundamental scale to that of the effective theory, 
there may be no good reason to choose a particular
relation among the charges.
On the other hand, if the gauge coupling is sufficiently small, then 
the renormalization-group running may be ignored. Furthermore, 
various theoretical developments within the last few years have 
demonstrated that the range of possibilities 
for physics at a fundamental scale is very wide, and hence
it is reasonable to consider charge assignments that are
different than those arising from traditional grand unified 
theories.

An example that is both simple and instructive is based on the 
$SU(3)_{C} \times SU(2)_{W} \times U(1)_Y \times U(1)_{B-L}$
gauge group. As we discussed in Section 4.1, in this case 
there is no tree-level mixing between the $Z$ and 
$Z^\prime \equiv Z_{B-L}$ bosons, because $z_H =0$. 
The best limits on the mass and coupling of the $Z_{B-L}$
boson is then set by the searches for direct $Z^\prime$
production in experiments at the Tevatron and LEPII.
For a $Z_{B-L}$ coupling to quarks and leptons of electromagnetic 
strength, the lower mass limit set by searches at CDF is around 
480 GeV. 
We reiterate though that
the gauge coupling is a free parameter that could be 
substantially smaller than the electromagnetic gauge 
coupling. If that is the case, then the Tevatron mass limits
no longer apply, and the best bounds for a $Z_{B-L}$ of mass
below 200 GeV are set at LEP. With the exception of a few narrow
mass intervals around the center-of-mass energies where the integrated 
luminosity is large, a $Z_{B-L}$ coupling to leptons as 
large as $10^{-2}$ would 
suffice to hide the narrow resonance from the LEP experiments.

For generic charge assignments with $z_H \neq 0$, the strongest  
bounds on the $Z^\prime$ parameters come from precision
measurements at the $Z$-pole. The presence of the new $Z'$ in general
induces at the tree-level both a shift in the $Z$ mass (expressed in terms
of the $T$ parameter) and a shift in its couplings from the standard model
values. For a $Z'$ coupling to fermions of typical electromagnetic strength,
we estimate the lower bound on the mass $M_{Z'}$ of the $Z'$ to be 
roughly in the $0.9 - 1.2$ TeV range.
As the $Z^\prime$ coupling to the Higgs doublet weakens, the lower bound on
$M_{Z'}$ from the indirect, $Z$-pole studies drops significantly (see Fig.~2).

We emphasize that we have studied so far only the ``tip of the iceberg''.
There are many avenues for research related to new gauge bosons.
It would be interesting to investigate systematically the 
possible charge assignments and limits on couplings when the
$U(1)_z$ charges are generation dependent (various examples
of this type have been analyzed recently in 
Ref.~\cite{Chivukula:2002ry}).
Furthermore, our simplifying assumption that there are no
``exotic'' fermions charged under the standard model group
could be dropped. Given that the scale of 
new physics is expected to be at the TeV scale, one could also 
consider an anomalous $U(1)_z$. 
Experimental limits on the parameter space associated with the
gauge bosons in each of these generalizations need to be derived.
In fact, only few dedicated searches for light narrow 
resonances have been performed, and therefore there is a possibility 
that the signal for a new gauge boson already exists in the current 
data.

\bigskip\bigskip

{\bf Acknowledgements}: We would like to thank Sekhar Chivukula, Stephen
Martin, Rabi Mohapatra, Maurizio Piai and German Valencia
for helpful comments.
This work was supported in part by grant DE-FG02-92ER-4074.
Fermilab is operated by University Research Association, Inc., under
contract DE-AC02-76CH03000.


\vfil
\end{document}